# Skyrmions with customized intensity distribution and trajectory

Yihan Tian[1], Guoxia Han[1,*], Shiru Song[1], Feiyang Zhang[1], Guangyi Wang[1], Qihui Zhao[1], Maoda Jing[1], Xianghua Yu[2]

1, College of Science, China University of Petroleum (East China), Qingdao 266580, Shandong, China
2, State Key Laboratory of Ultrafast Optical Science and Technology, Xi'an Institute of Optics and Precision Mechanics, Chinese Academy of Sciences, Xi'an, Shaanxi, China

**\*Corresponding author:** E-mail address：**gxhan@upc.edu.cn**   (G. Han)

**Abstract:** Optical skyrmions, which are topological protection quasi-particles with nontrivial textures, hold a pivotal focus in current structured light research for their potential in diverse applications. In this work, the angular spectrum theory is first introduced into the generation of optical skyrmions and modulation of the intensity and trajectory of skyrmions at will. We propose a novel theoretical approach for the generation of skyrmions, including Néel-type, Bloch-type, anti-type and $2^{nd}$ order. The simultaneous and independent modulation of intensity distribution and trajectory of isolated skyrmions is first achieved with the combination of phase-shifting theory with angular spectrum theory. By controlling the displacement phase factor (DPF), the customized shape of skyrmions array with controllable intensity distribution and trajectory is also generated. Our findings in this work allow a greater exploration of skyrmions, which promote applications in particle manipulation and high-density storage.

# 1 Introduction

Skyrmions, as topologically protected quasi-particles possessing non-trivial textures, have received escalating interests [1]. The concept of skyrmions was initially proposed as a topological soliton solution by Tony Skyrme in 1962 [2, 3]. Later, the emerging quasi-particle has been demonstrated in many systems, such as liquid crystals [4], Bose-Einstein condensates [5] and quantum hall systems [6]. Specifically in condensed matter systems, magnetic skyrmions have garnered significant scientific attention owing to its unique properties, such as nanoscale size, topologically protected stability, and ultralow driving current densities [7], [8]. Great achievements of the research on magnetic skyrmions have been promoted and applied in many fields, including information storage [9], [10], logical operations [11], neuromorphic computing [12], et.al.

As the counterpart of magnetic skyrmions, optical skyrmions were discovered by Yuan et al. in 2018 within the evanescent electric field using the experiment of small-angle neutron scattering (SANS) [13]. Since then, optical skyrmions have performed as a frontier research topic in the optical domain because of the same topologically protected properties similar to their magnetic counterparts [14]-[31]. Subsequently, many studies have realized and modulated optical skyrmions in diversified 3D vector fields, including electromagnetic field vectors [14]-[16], spin vectors [17]-[19], Stokes vectors [20]-[28], Poynting vectors [29], pseudo spin vectors [30] and Mie scattering fields[31]. Characterized by the pivotal attributes of propagation-stable in free space, Stokes vector skyrmions have drawn more and more attention in recent years. The predominant approach to constructing isolated stokes skyrmions employs the coherent superposition of a pair of orthogonally polarized Laguerre-Gaussian beams [20]-[23] or Bessel beams[25], [26]. However, recent studies with this approach for the generation of isolated skyrmions are constrained to propagating along linear trajectories. Parallel to the studies of isolated skyrmions, many researchers focused on morphological versatility of skyrmions with recent demonstrations covering skyrmion lattices and arrays. Yuan et al. created Stokes skyrmions array with different pattern (such as letters "L" and "Y") by controlling the wavevectors of tightly focusing azimuthally polarized vortex beam [32]. Shen et al. imprinted cubic phase modulation onto the conventional skyrmionic beams to realize self-accelerating skyrmion lattices [33]. But in this condition, the resulting parabolic trajectory comes from the imposed phase similar to self-accelerating Airy beams, making the method fundamentally incapable of realizing arbitrary paths. To the best of our knowledge, the present isolated and skyrmions lattices and arrays in most studies exhibit strictly straight or parabolic paths. Moreover, the modulation of the intensity distribution of skyrmions along the propagation direction holds critical significance for skyrmions applications, yet it has not been investigated until now. So, achieving simultaneous and independent on-demand control over both the trajectory and the intensity distribution of individual and arrayed skyrmions remains a critical unmet challenge.

In this work, we pioneer the introduction of angular spectrum theory into the research of skyrmions and a novel method is proposed to realize the on-demand modulation of intensity distribution and trajectory of skyrmions. Utilizing the angular spectrum function of Bessel beams, we construct diverse topological textures such as Néel-type, Bloch-type, anti-type and $2^{nd}$ order skyrmions. By embedding regulation functions in the angular spectrum, the intensity distribution and trajectory of isolated skyrmions can be simultaneously and independently modulated. Moreover, the customized shape of skyrmions array with controllable intensity distribution and trajectory is achieved by controlling the displacement phase factor (DPF). Our finding has significance in particle manipulation and high-density storage, opening new directions in optical Stokes skyrmions research.

## 2 Methods

### 2.1 Construction of skyrmions in terms of the angular spectrum theory

Optical skyrmions, with non-trivial texture in vectorial fields, are characterized by topological invariant Q (skyrmion number) which quantify the wrapping of the polarization field around the vector sphere. The skyrmion number Q of the skyrmions constructed in a Stokes vector field within the cylindrical coordinates ($r, \phi, z$) is

defined by Eq. 1 [13], [34]:

$$Q = \frac{1}{4\pi} \iint_D \vec{S} \cdot (\partial_r \vec{S} \times \partial_\phi \vec{S}) r dr d\phi \tag{1}$$

where $S = (S_1, S_2, S_3)/S_0$ is the normalized Stokes vector, where $S_0$, $S_1$, $S_2$ and $S_3$ represent the Stokes parameters for the polarization state of beams, respectively. D is the integral area.

The skyrmions in a Stokes vector field can be constructed by the superposition of a pair of orthogonally polarized Bessel beams, as follows [25]:

$$E(r, \phi, z) = \frac{1}{\sqrt{2}}[B_{l_1}(r, \phi, z)|R\rangle + e^{i\theta_0} B_{l_2}(r, \phi, z)|L\rangle] \tag{2}$$

where $|R\rangle$ and $|L\rangle$ represent the right-handed and left-handed circular polarization state, respectively. $B_{l_1}$ and $B_{l_2}$ are the $l_1^{th}$ order and the $l_2^{th}$ order Bessel beams, and the phase difference between them is denoted by $\theta_0$. The expression of any order Bessel beam is as follows:

$$B_l(r, \phi, z) = J_l(k_r r) e^{i(l\phi + k_z z)} \tag{3}$$

where $J_l(\cdot)$ represents the Bessel function of the first kind, $k_z$ and $k_r = \sqrt{k^2 - k_z^2}$ are the axial and radial components of the wave-vector $k = \frac{2\pi}{\lambda}$, respectively, where $\lambda$ is the wavelength of the beam.

The field distribution of a monochromatic light field propagating along the z-axis in free space can be expressed as the superposition of a series of plane wave components with different frequencies. This representation can be mathematically interpreted as the transformation of the spatial distribution of the light field into the frequency domain (i.e. angular spectral) through a Fourier transform. Correspondingly, the relationship of Bessel beam $B_l(r, \phi, z)$ propagating along z-axis and its angular spectrum can be described by a Fourier transform as [35], [36]:

$$B_l(r, \phi, z) = (\frac{1}{2\pi})^2 \int_0^{2\pi} \int_{-k}^{k} A(k_r, \varphi) e^{ik_r r \cos(\varphi - \phi)} e^{ik_z z} k_r dk_r d\varphi \tag{4}$$

where $(k_r, \varphi, k_z)$ represent the three-dimensional cylindrical coordinates in the Fourier-space (k-space) and $A(k_r, \varphi)$ is the angular spectrum function.

Considering the axisymmetry property of Bessel beams and utilizing the integral formula of Bessel functions $J_l(k_r r)e^{il\phi} = \frac{i^{-l}}{2\pi}\int_0^{2\pi} e^{ik_r r\cos(\varphi-\phi)} e^{il\varphi} d\varphi$, the expression of angular spectrum function represented by Equation 3 can be simplified as:

$$B_l(r, z) = \frac{(-i)^{-l}}{2\pi}\int_{-k}^{k} A(k_r) J_l(k_r r) e^{il\phi} e^{ik_z z} k_r\, dk_r \qquad (5)$$

Given the special energy distribution of Bessel beams, we merely pay attention to the complex amplitude distribution on the their vortex ring with radius $r_m$, where the expression of Bessel beam can be given by:

$$B_l(r=r_m, z) = C_m J_l(k_{r_m} r_m) e^{il\phi} e^{ik_{zm}z} = e^{il\phi} e^{ik_{zm}z} \qquad (6)$$

in which, $C_m$ is the normalization constant and $r_m$ represents the radius of the vortex ring of Bessel beam.

By substituting Equation (5) into Equation (4) and using a inverse Fourier transform, the angular spectrum function of the $l^{th}$ order Bessel beam can be described as:

$$A_l(\sqrt{k^2-k_z^2}, \varphi) = \frac{2\pi e^{il\varphi}\int_{-\infty}^{\infty} e^{ik_{zm}z} e^{-ik_z z} dz}{(-i)^{-l} rect(\frac{k_z}{2k}) J_l(\sqrt{k^2-k_z^2} r_m) k_z} \qquad (7)$$

where rect(·) represent the rectangle function and $k_z$ is truncated within a constrained range (-$k$, $k$). According to Equation 1, the skyrmions can be constructed by the superposition of a pair of orthogonally polarized Bessel beams. Therefore, the generation of stokes vector skyrmions in the framework of angular spectrum theory can be achieved with the following equation:

$$E(r, \phi, z) = \frac{1}{\sqrt{2}}[\mathcal{F}^{-1}(A_{l_1}(\sqrt{k^2-k_z^2}, \varphi))|R\rangle + e^{i\theta_0}\mathcal{F}^{-1}(A_{l_2}(\sqrt{k^2-k_z^2}, \varphi))|L\rangle]$$
$$= \frac{1}{\sqrt{2}}[U_{l_1}(r, \phi, z)|R\rangle + e^{i\theta_0} U_{l_2}(r, \phi, z)|L\rangle] \qquad (8)$$

where $\mathcal{F}^{-1}(\cdot)$ represents the inverse Fourier transform and $U_l(\cdot)$ is $l^{th}$ order Bessel beam constructed by the angular spectrum theory. By tuning the basic parameters $\theta_0$ and $\Delta l = l_2 - l_1$, the skyrmions of different type can be generated.

## 2.2 Construction of skyrmions with customized intensity distribution and trajectory

As we all know that the angular spectrum can govern the propagation direction and intensity of plane wave components of optical field, while the superposition and interference of these components ultimately determine the morphological characteristics of the spatial field. Therefore, the precise regulation of intensity distribution and trajectory of skyrmions can be achieved by manipulating the angular spectrum function of the beams.

The integration in the angular spectrum function in Equation (7) can be mathematically interpreted as a summation of infinitesimally thin slice according to the idea of calculus. If each thin slice is tailored in the angular spectrum domain and the modified spectral components are accumulated, the on-demand skyrmions can be obtained. In the tiny range of $(z, z + \Delta z)$, the angular spectrum function is expressed as:

$$A_z(\sqrt{k^2 - k_z^2}, \varphi) = \frac{2e^{il\varphi}e^{ik_{zm}z}e^{-ik_z z}}{2\pi i^l rect(\frac{k_z}{2k})J_l(\sqrt{k^2-k_z^2}r_m)k_z} \tag{9}$$

According to the homogeneity principle of Fourier transform, multiplying the angular spectrum function A(k) by a scaling factor α in the frequency domain exhibits exact equivalence to synchronously scaling its spatial-domain counterpart by the identical coefficient α. Correspondingly, for a specific axial position, if an amplitude factor is applied to the tiny slice, the intensity distribution of skyrmions can be modulated. This amplitude modulation factor is defined as the Amplitude Compensation Function *I* (*z*) (ACF), which varies as a function of the z-axis. So, the angular spectrum function of the tiny slice with the capability of intensity distribution regulation can be written as:

$$A_z(\sqrt{k^2 - k_z^2}, \varphi) = \frac{2e^{il\varphi}\sqrt{I(z)}e^{ik_{zm}z}e^{-ik_z z}}{2\pi i^l rect(\frac{k_z}{2k})J_l(\sqrt{k^2-k_z^2}r_m)k_z} \tag{10}$$

By accumulating the angular spectrum function of the tiny slices, on-demand control of the intensity of skyrmions along the propagation direction can be achieved.

Additionally, according to Fourier phase-shifting theory, a complex phase factor $e^{-ik_x x_0 + ik_y y_0}$ multiplied to $A(k_x, k_y)$ in the frequency domain will enable a corresponding spatial lateral displacement $(x_0, y_0)$ in the optical field distribution, as mathematically expressed as:

$$U(x - x_0, y - y_0) = \mathcal{F}^{-1}\{A(k_x, k_y) \cdot e^{-ik_x x_0 + ik_y y_0}\} \tag{11}$$

where $U(x, y)$ and $A(k_x, k_y)$ represent the spatial distribution and frequency distribution, respectively. $k_x = k_r cos\varphi$ and $k_y = k_r sin\varphi$ are the radial wave-vector in the Cartesian coordinates and $e^{-ik_x x_0 + ik_y y_0}$ is the phase factor in the frequency domain, where $x_0$ and $y_0$ are lateral displacement along the x and y directions in the spatial domain, respectively.

Similar to the intensity distribution regulation above, for a specific axial position, axial dependency functions *f(z)* and *g(z)* are introduced into the angular spectrum function to regulate the trajectory of skyrmions. *f(z)* and *g(z)* are collectively termed the trajectory modulation function (TMF). Further integrating the TMF into the angular spectrum function of the tiny slice with the capability of regulating intensity, we can obtain the angular spectrum function of the tiny slice with controlled intensity distribution and trajectory, as follows:

$$A_z(\sqrt{k^2 - k_z^2}, \varphi) = \frac{2e^{il\varphi}\sqrt{I(z)}e^{ik_x f(z) + ik_y g(z) + ik_z m z} e^{-ik_z z}}{2\pi i^l rect(\frac{k_z}{2k}) J_l(\sqrt{k^2 - k_z^2} r_m) k_z} \tag{12}$$

Stacking the modulated angular spectrum function of the tiny slice, we propose the angular spectrum function of a customized Bessel beam, featuring tailored intensity distribution and trajectory:

$$A'_l(\sqrt{k^2 - k_z^2}, \varphi) = \frac{e^{il\varphi} \int_{-\infty}^{\infty} \sqrt{I(z)} e^{-ik_x f(z) - ik_y g(z) + ik_z m z} e^{-ik_z z} dz}{2\pi i^l rect(\frac{k_z}{2k}) J_l(\sqrt{k^2 - k_z^2} r_m) k_z} \tag{13}$$

Consequently, according to Eq.2, the stokes vector skyrmions with on-demand intensity distribution and trajectory in the framework of angular spectrum theory can be obtained as follows:

$$E'(r, \phi, z) = \frac{1}{\sqrt{2}}[\mathcal{F}^{-1}(A'_{l_1}(\sqrt{k^2 - k_z^2}, \varphi))|R\rangle + e^{i\theta_0} \mathcal{F}^{-1}(A'_{l_2}(\sqrt{k^2 - k_z^2}, \varphi))|L\rangle]$$

$$= \frac{1}{\sqrt{2}} [U'_{l_1}(r, \phi, z)|R\rangle + e^{i\theta_0} U'_{l_2}(r, \phi, z)|L\rangle] \quad (14)$$

where $U'_l(\cdot)$ is $l^{th}$ order Bessel beam with controllable intensity distribution and arbitrary trajectory. The desired skyrmions can be synchronously and independently regulated with angular spectrum control functions ACF and TMF, and basic parameters $\Delta l$ and $\theta_0$. ACF controls the intensity distribution along the propagation axis of the skyrmions, while TMF, comprising $f(z)$ and $g(z)$, are employed to control the displacement in the x and y directions, with regard to the z position, to achieve the modulation of trajectory, respectively. Precisely engineered the phase difference $\theta_0$ and order difference $\Delta l = l_2 - l_1$ enable the creation of skyrmions with different types.

In the preliminary work, precise control over trajectory of skyrmions is achieved by deconstructing the angular spectrum function into discrete thin slices and applying Fourier phase-shifting theory independently to each slice. Further, we directly extend this theory to manipulation of the complete angular spectrum function. The overall shifting of isolated skyrmions is achieved through multiplying a displacement phase factor $e^{[-i(k_x \Delta x_i + k_y \Delta y_i + k_z \Delta z_i)]}$ (DFP) in modulated angular spectrum function of Bessel beams. By coherent superposition of the isolated skyrmions at specific locations, skyrmions array with customized intensity distribution, trajectory and shape can be realized. The representation of a skyrmion array composed of n skyrmions can be derived as follows:

$$E_a(r, \phi, z) = \frac{1}{\sqrt{2}} \sum_{i=0}^{n} [\mathcal{F}^{-1}(A^i_{l_i}(\cdot) \cdot \varepsilon_i)|R\rangle + e^{i\theta_0} \mathcal{F}^{-1}(A^i_{l_{i+1}}(\cdot) \cdot \varepsilon_i)|L\rangle]$$

$$= \frac{1}{\sqrt{2}} \sum_{i=0}^{n} [U^i_{l_i}(\cdot)|R\rangle + e^{i\theta_0} U^i_{l_{i+1}}(\cdot)|L\rangle] \quad (15)$$

where $\varepsilon_i = e^{[-i(k_x \Delta x_i + k_y \Delta y_i + k_z \Delta z_i)]}$ is the displacement phase factor in the frequency domain and $(\Delta x_i, \Delta y_i, \Delta z_i)$ represents the spatial shift of the $i^{th}$ isolated skyrmion. The displacement phase factor $\varepsilon_i$ hold the spatial shift of isolated skyrmions to customize the shapes of skyrmions array. The skyrmions in the array can be synchronously and independently controlled by angular spectrum control functions ACF and TMF and basic parameters $\Delta l$ and $\theta_0$, as in the previous method of regulating isolated skyrmions.

## 3 Results and discussion

Based on the theoretical framework established in section 2.1, we realize the generation of skyrmions in Bessel modes utilizing angular spectrum theory and the Néel-type, Bloch-type, anti-type and $2^{nd}$ order skyrmions are depicted in Figure 1. By employing Equation 7, the required angular spectrum distributions of Bessel beams with topological charges $l$ = 0, 1, -1, 2 are calculated. The corresponding results are shown in Figure 1(a), where the luminance corresponds to the amplitude and the color-coding represents the phase. Subsequently, we substitute the angular spectrum function into Equation 8 to obtain the desired skyrmion, with the specific basic parameters used in the simulations detailed in Table 1 (below).

Table 1. The basic parameters of Néel-type, Bloch-type, anti-type and $2^{nd}$ order skyrmions

| $l_1$ | $l_2$ | $\theta_0$ | type | Q |
|---|---|---|---|---|
| 0 | 1 | 0 | Néel | 1 |
| 0 | 1 | $\frac{\pi}{2}$ | Bloch | 1 |
| 0 | -1 | 0 | Anti | -1 |
| 0 | 2 | 0 | $2^{nd}$ order | 2 |

Figure 1(b) and (c) present the polarization distribution and 3D view of the Stokes vector of the Néel-type, Bloch-type, anti-type and $2^{nd}$ order skyrmions, respectively. The right-handed and left-handed polarization states are represented by red and blue ellipses in Figure 1(b), which correspond to upward and downward arrows in the Stokes vector shown in Figure 1(c), respectively. Here, to clearly illustrate the texture of the skyrmion center, only the Stokes vector components within the first azimuthal period of the polarization structure are presented. The Néel-type skyrmion exhibits hedgehog textures and the Bloch-type is vortex configurations, with interconversion enabled by the adjustment of the parameter $\theta_0$. Concurrently, by keeping the topological charge parameters $l_1 = 0$ and setting $l_2 = -1$ and 2, the Anti-type with saddle texture and *2nd* order skyrmion are obtained.

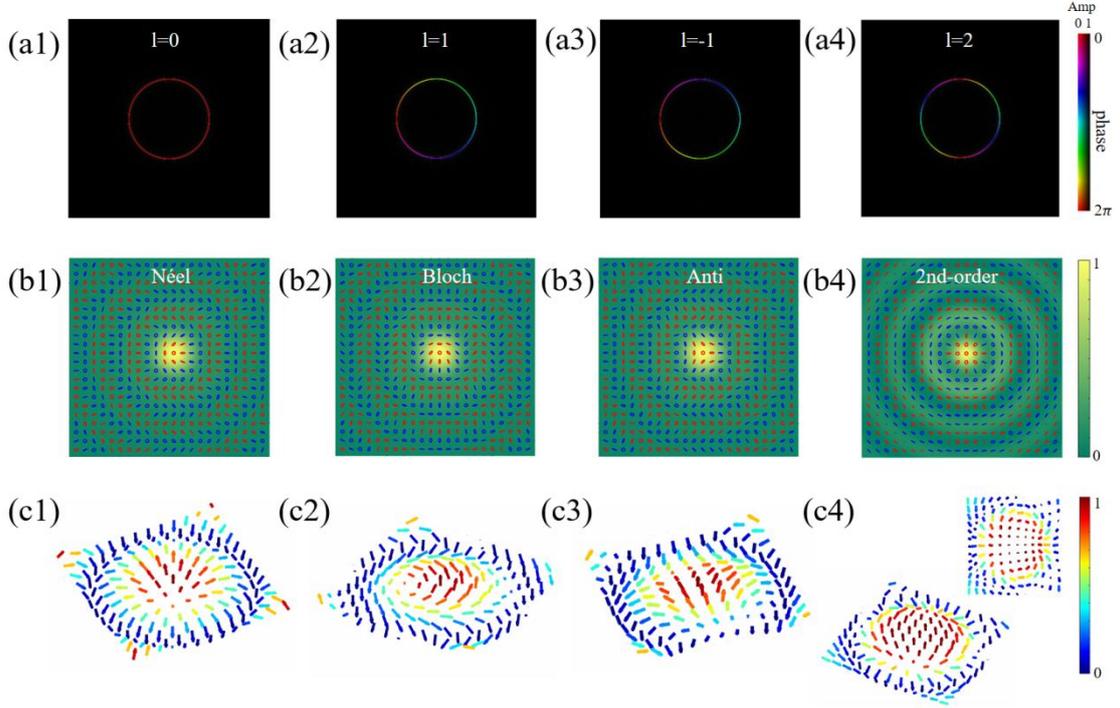

Figure 1: Optical skyrmions generated based on angular spectrum expression of Bessel beams. (a) The angular spectrum distribution of $l^{th}$ order Bessel beams. (b) The transverse polarization profile of optical skyrmion textures: Néel-type, Bloch-type, anti-type and *2nd* order skyrmions from left to right. (c) 3D view of the Stokes vector corresponding to the polarization distribution in (b).

### 3.1 Isolated skyrmions

Building upon the generation of skyrmions with different types, the ACF and TMF were introduced into angular spectrum function of Bessel beams to modulate the intensity distribution and trajectory of isolated skyrmions. We take the Néel-type as an example to explore the specific influence of these two modulating functions on the characteristics of skyrmions. Firstly, we achieve the trajectory control of skyrmions while maintaining the uniform-like intensity distribution and Figure 2 present the corresponding result. The functions ACF and TMF are set as I(z) =1, $f(z) = 0.15cos[2\pi(z/f + 1)]$ and g(z) = 0, respectively. Utilizing Equation (13), the modulated angular spectrum distribution of $l = 0^{th}$ and $1^{st}$ order Bessel beams are calculated and illustrated in Figure 2(a). By setting $l_1 = 0$, $l_2 = 1$ and $\theta_0 = 0$, the corresponding curved Néel-type skyrmion can be realized utilizing the established Equation (14). Figure 2(b) and (c) depict the longitudinal intensity distribution of the desired skyrmion in the x-z plane and the corresponding normalization intensity profile of the beam along propagation direction, respectively. It is evident that the electric field exhibits a parabola trajectory and uniform intensity distribution. The simulation results show close agreement with the theoretical assumptions of the ACF and TMF. To investigate the topological protection characteristics of the modulated skyrmion, four randomly spatial positions are chosen to visualize the polarization and Stokes vector distribution of transverse profiles, as shown in Figure 2(d1-d4) and (e1-e4), the transverse profiles of the isolated skyrmion at z =− 0.075m , 0m , 0.075m and 0.14m given in the order they occurred in the propagation. It is worth noting that the position of the main lobe of the beam alters in the x-y plane, however, the polarization ellipse distribution and Stokes vector reveal that the skyrmion still maintains the Néel-type topological structure. These results indicate that the topological protection characteristics of isolated skyrmion along serpentine trajectory are preserved.

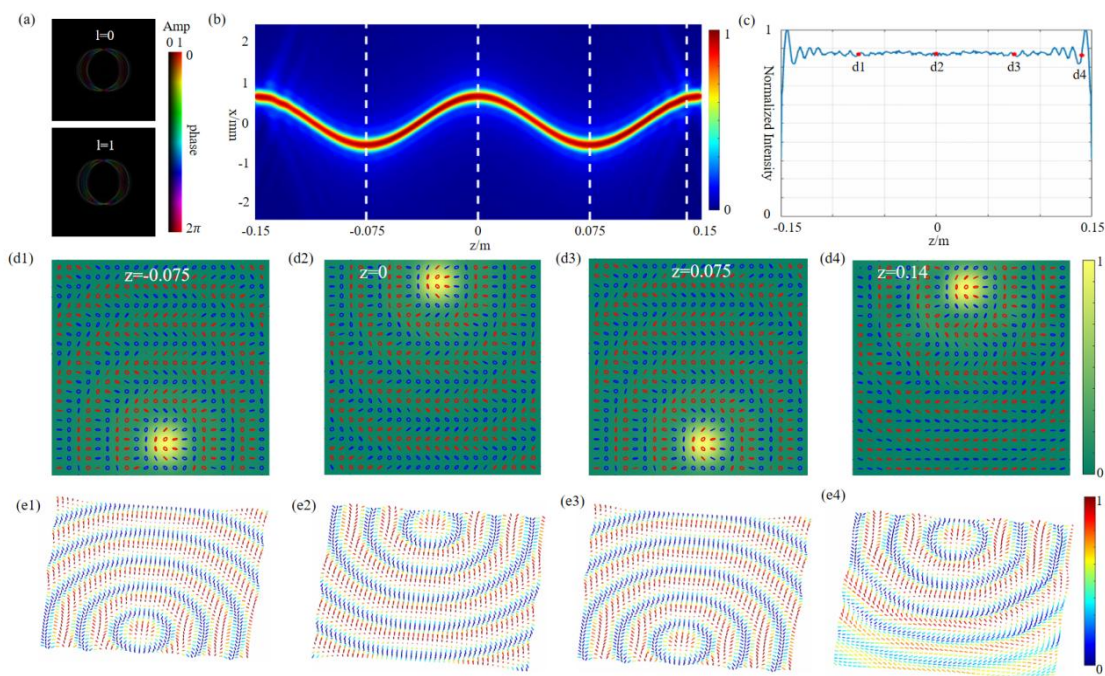

Figure 2: Isolated skyrmion propagating along a serpentine trajectory. (a) The angular spectrum of l = 0$^{th}$ and 1$^{st}$ order Bessel beams. (b) Spatial-intensity profile in the x-z plane. (c)1D normalized

intensity distribution. (e1-e4) Polarization distribution at distinct propagation planes: $z = -0.075m$, $0m$, $0.075m$ and $0.14m$. (f1-f4) 3D view of the Stokes vector corresponding to the polarization distribution in (b).

Secondly, we achieve the intensity distribution and trajectory control of skyrmions and Figure 3 present the corresponding result. The parameters are set as $I(z) = \cos(2\pi(\frac{z}{f} + 1))$, $f(z) = 0.15 \times 10^{-3}(z/f)^2 - 0.05 \times 10^{-3}$ and $g(z) = 0$. Figure 3(a) illustrates the angular spectrum distribution of Bessel beams with orders $l = 0^{th}$ and $1^{st}$, determined by the reset parameters ACF and TMF. The longitudinal cross-sectional view and normalization intensity distribution of the modulated skyrmion are depicted as Figure 3(b) and (c), respectively. We can see that in this situation, the electric field holds a parabola trajectory while periodic intensity distribution is achieved, similar to a "curved chain", with the simulation results maintaining a high degree of consistency with the theoretical assumptions of the ACF and TMF. The polarization distribution and Stokes vector at four discrete cross-sectional planes: z = 0m, 0.055m, 0.095m and 0.14m are separately depicted in Figure 3 (d1-d4) and (e1-e4). It is clear that the topological structure of the "curved chain" skyrmion still exhibits the Néel-type texture, confirming that our method based on angular spectrum not only ensures robust control over intensity and trajectory, but also maintains the topological stability of skyrmions. These results indicate that the intensity and trajectory of isolated skyrmions can be synchronously and independently modulated with our method and the topological protection characteristics of modulated isolated skyrmion are preserved.

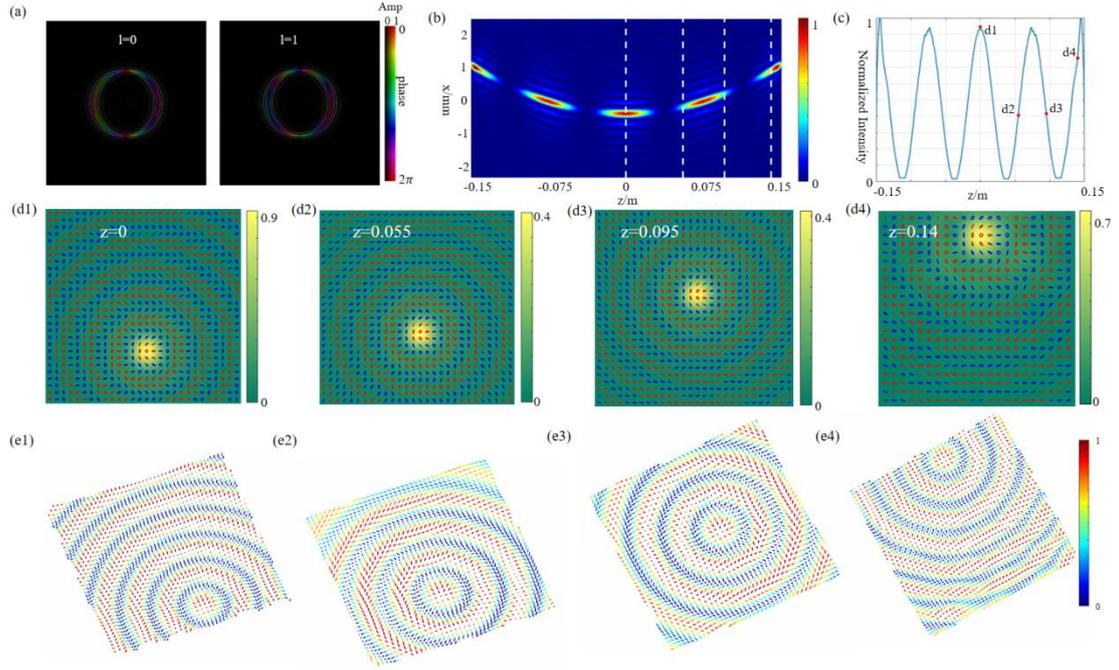

Figure 3: Skyrmion with cosine intensity distribution and a parabola trajectory. (a) The angular spectrum of l = 0$^{th}$ and 1$^{st}$ order Bessel beams. (b) Intensity map in the x-z plane over a propagation range of z=-0.15m to z=0.15m. (c) Longitudinal intensity evolution of the skyrmion over a propagation range. (d1-d3) Polarization distribution at distinct propagation planes: z=0m, 0.055m, 0.095m, 0.14m. (e1-e3) 3D view of the Stokes vector corresponding to the polarization distribution in (d).

### 3.2 Skyrmions arrays

Based on the above research, a versatile family of isolated skyrmions with on-demand tailored intensity distribution and trajectory can be generated by tuning the functions $I(z)$, $f(z)$ and $g(z)$. In this section, skyrmions array with on-demand intensity distribution, trajectory and shape are generated and modulated in our proposed framework. The custom-patterned skyrmions array can be generated through the coherent superposition of isolated skyrmions at specific spatial locations, which are dynamically steered by modulating the parameters $\Delta x, \Delta y$ and $\Delta z$ in the displacement phase factor. Crucially, this framework also enables the independent and synchronous control of each skyrmion of intensity distribution and trajectory in the array, achieved by dynamically adjusting their respective functions I(z), f(z), g(z) as detailed in the previous section.

Figure 4 depicts a square-type skyrmions array formed by four Néel-type skyrmions with linear intensity distribution, where the parameters are set to I(z)= 0.5( − z/f + 1) , f(z) = 0 and g(z) = 0 . The modulated angular spectrum distribution of $l = 0^{th}$ and $1^{st}$ order Bessel beams with line intensity distribution are illustrated in Figure 4(a). The specific position parameters, displacement parameters ($\Delta x_i, \Delta y_i, \Delta z_i$) of these four skyrmions are set as (8×10⁻⁴, -8×10⁻⁴, 0), (8×10⁻⁴, 8×10⁻⁴, 0), (-8×10⁻⁴, -8×10⁻⁴, 0), (-8×10⁻⁴, 8×10⁻⁴, 0), respectively. The 3D oblique sectional view of square-arranged skyrmions array is presented in Figure 4(b) with Figure 4(c) depicting the transverse view. Figure 4(d) presents the normalized intensity distribution of each skyrmion, revealing that each skyrmion constituent is monotonic linear intensity distribution along its trajectory. Similar to isolated skyrmions, we also illustrate the polarization distribution and Stokes vector of each skyrmion in the array at z=0.14m in Figure 4(e1-e4) and (f1-f4). It can be seen that the four units all show Néel-type skyrmion texture. The results demonstrate that the skyrmions array can be achieved by our proposed method in the framework of the angular spectrum theory.

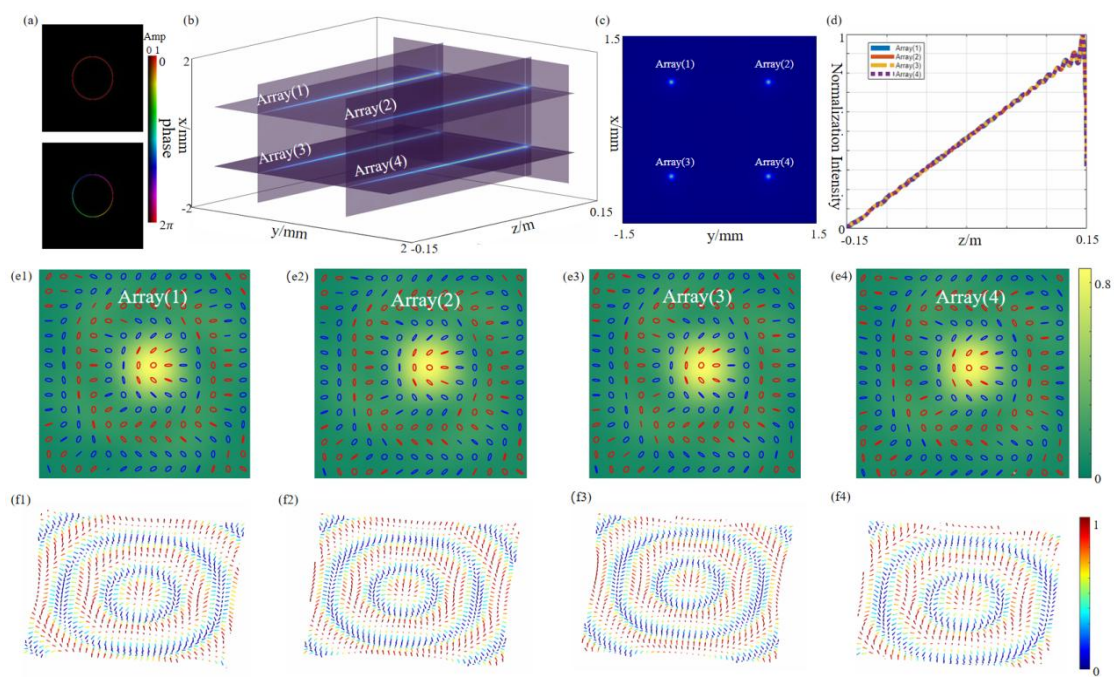

Figure 4: The square-type skyrmions array composed of four isolated skyrmions of identical characteristics. (a) The angular spectra distribution utilized to construct array. (b) 3D oblique sectional view of the array. (c) Cross-sectional view at z=0.14m. (d) 1D normalized intensity

distribution.(c) Cross-sectional view at z=0.14m. (e1)-(e5) Polarization distribution of isolated skyrmions within the array at the cross-section at z=0.14m. (f1)-(f5) 3D view of the Stokes vector corresponding to the polarization distribution in (e).

In addition to the formation of a square-type skyrmions array consisting of four identical skyrmions with line intensity distribution, an L-type skyrmions array comprising five skyrmions that are synchronously and independently regulated is also presented, and the results are shown in Figure 5. The key parameters governing the synchronous and independent regulation of each skyrmion in the array are provided in Table 2, including the displacement parameters ( $\Delta x_i, \Delta y_i, \Delta z_i$ ), angular spectrum control functions ( $I(z), f(z)$ and $g(z)$ ) and basic parameters ($l_i$, $l_{i+1}$ and $\theta_0$ ). The angular spectrum distribution required to construct each skyrmion within the arrays are calculated according to Equation (13) utilizing the angular spectrum control parameters, with the corresponding results presented in Figure 5(a). The displacement phase factors, determined by the displacement parameters, are imposed onto this calculated angular spectrum distribution. Employing Equation (15), the L-type skyrmions array is constructed through the superposition of isolated skyrmions with specific spatial displacement. Figure 5(b) and (d) depict the 3D oblique view of the L-type skyrmion array and the normalized intensity distribution of each skyrmion, respectively. The distinct propagation behaviors of these five skyrmion units are clearly observed, which exhibit a linear intensity distribution with linear trajectory, uniform intensity distribution with parabolic trajectory, a cosine intensity distribution with parabolic trajectory, a cosine intensity distribution with linear trajectory and uniform intensity distribution with serpentine trajectory, respectively. Figure 5(d)-(f) show the transverse distribution, polarization distribution and Stokes vector of this L-shaped skyrmion array at z=0.14m, where the types of skyrmion are Néel-type texture, anti-type texture, high-order texture, anti-type texture and Bloch-type texture in turn. This results confirm that our method of achieving skyrmion arrays constructed based on angular spectrum not only ensures robust control over the shape of array, but also modulate the intensity distribution and trajectory of each skyrmion within the array.

Table 2. The parameters of L-type skyrmions array

| Array | basic parameters | | | angular spectrum control functions | | | displacement parameters |
|---|---|---|---|---|---|---|---|
| | $l_i$ | $l_{i+1}$ | $\theta_0$ | $I(z)$ | $f(z)$ | $g(z)$ | $(\Delta x_i, \Delta y_i, \Delta z_i)$ |
| (1) | 0 | 1 | 0 | $0.5(-\frac{z}{f}+1)$ | 0 | 0 | (s, -s, 0) |
| (2) | 0 | -1 | 0 | 1 | $0.05\times 10^{-3}\left[(\frac{z}{f})^2-1\right]$ | 0 | (0, -s, 0) |
| (3) | 0 | 2 | 0 | $cos(2\pi(\frac{z}{f}+1))^2$ | $0.05\times 10^{-3}\left[(\frac{z}{f})^2-1\right]$ | 0 | (-s, -s, 0) |
| (4) | 0 | -1 | 0 | $cos(2\pi(\frac{z}{f}+1))^2$ | 0 | 0 | (-s, 0, 0) |
| (5) | 0 | 1 | $\frac{\pi}{2}$ | 1 | $0.06\times 10^{-3}cos(2\pi(\frac{z}{f}+1))$ | 0 | (-s, s, 0) |

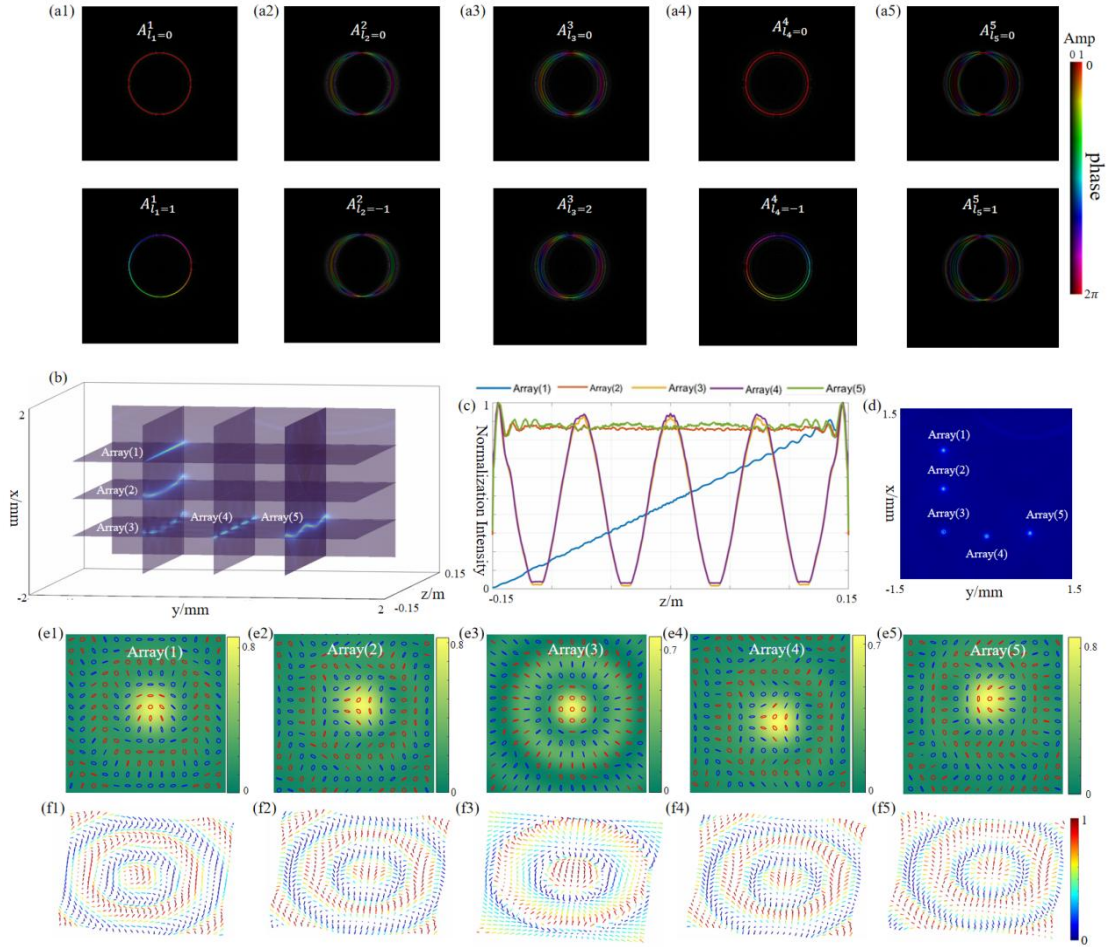

Figure 5: The L-type skyrmions array composed of five isolated skyrmions of distinct characteristics. (a)The angular spectrum distributions utilized to construct array. (b) 3D oblique sectional view of the array. (c)1D normalized intensity distribution.(d) Cross-sectional view at

z=0.14m. (e1)-(e5) Polarization distribution of isolated skyrmions within the array at the cross-section at z=0.14m. (f1)-(f5) 3D view of the Stokes vector corresponding to the polarization distribution in (e).

## 4 Conclusion

In conclusion, we have demonstrated the full control of the intensity distribution and trajectory of isolated skyrmions and skyrmions array in Bessel modes in the framework of angular spectrum theory. Firstly, we pioneer the introduction of angular spectrum theory into the research of skyrmions and provide a new theoretical approach for the generation of skyrmions with different types (such as Néel-type, Bloch-type, anti-type and $2^{nd}$ type). Additionally, based on the presented method and combined with the homogeneity principle and phase-shifting theory of Fourier transform, the intensity and trajectory of skyrmions are modulated on-demand synchronously and independently. Furthermore, we extend the study of complex isolated skyrmions to array and achieve the generation of the custom-patterned skyrmions array with intricate intensity distribution and trajectory. The proposed method significantly increases the degrees of freedom in controlling the skyrmions, offering potential applications in studying particle manipulation, as well as providing theoretical reference and support for further research of skyrmions.

**Acknowledgments**

This work was supported by National Key Research and Development Program of China (2023YFF1205704) and Western Youth Scholars Project of Chinese Academy of Sciences (No. XAB2022YN13). The data presented in this study were generated independently and were not funded by the RCF.

**Conflict of interests**

The authors declare that they have no known competing financial interests or personal relationships that could have appeared to influence the work reported in this paper.